\documentclass[prd,twocolumn,reprint,preprintnumbers,nofootinbib,superscriptaddress]{revtex4-1}
\pdfoutput=1
\usepackage{cancel}

\input{header}

\begin{document}
\title{Search for Charm-quark Production via Dimuons in Neutrino Telescopes}

\author{ChuanLe Sun}
\email{chlsun60@sjtu.edu.cn}
\affiliation{INPAC, Shanghai Key Laboratory for Particle Physics and Cosmology, School of
Physics and Astronomy, Shanghai Jiao Tong University, Shanghai 200240, China}

\author{Fuyudi Zhang}
\email{zhangfydphy@sjtu.edu.cn}
\affiliation{Tsung-Dao Lee Institute, Shanghai Jiao Tong University, Shanghai 200240, China}

\author{Fan Hu}
\email{fan_hu@pku.edu.cn}
\affiliation{Department of Astronomy, School of Physics, Peking University, Beĳing 100871, China}

\author{Donglian Xu}
\email{donglianxu@sjtu.edu.cn}
\affiliation{Tsung-Dao Lee Institute, Shanghai Jiao Tong University, Shanghai 200240, China}
\affiliation{INPAC, Shanghai Key Laboratory for Particle Physics and Cosmology, School of
Physics and Astronomy, Shanghai Jiao Tong University, Shanghai 200240, China}

\author{Jun Gao}
\email{jung49@sjtu.edu.cn}
\affiliation{INPAC, Shanghai Key Laboratory for Particle Physics and Cosmology, School of
Physics and Astronomy, Shanghai Jiao Tong University, Shanghai 200240, China}
\affiliation{Key Laboratory for Particle Astrophysics and Cosmology (MOE), Shanghai 200240, China}
\affiliation{Center for High Energy Physics, Peking University, Beijing 100871, China}

\begin{abstract}
Dimuon events induced by charm-quark productions from neutrino deep inelastic scattering (DIS) processes have been studied in traditional DIS experiments for decades. 
The recent progress in neutrino telescopes makes it possible to search such dimuon events at energies far beyond laboratory scale. 
In this paper, we construct a simulation framework to calculate yields and distributions of dimuon signals in an IceCube-like km$^3$ scale neutrino telescope.
Due to experimental limitation in the resolution of double-track lateral distance, only dimuon produced outside the detector volume are considered.
Detailed information about simulation results for ten years exposure is demonstrated.
Both an earlier work~\cite{Zhou:2021xuh} and our work study a similar situation, we therefore use that paper as a baseline to conduct comparisons.
We then estimate the impacts of different calculation methods of muon energy losses.
Finally, we study the experimental potential of dimuon searches under the hypothesis of single-muon background-only. 
Our results based on a simplified double-track reconstruction indicate a moderate sensitivity especially with the ORCA configuration.
Further developments on both the reconstruction algorithm and possible detector designs are thus required, and are under investigation.
\end{abstract}

\maketitle

\section{Introduction}\label{sec:introduction}
The weakness of interactions with matter renders neutrinos ideal messengers to their sources. 
They are mostly produced from the decay of secondary pions and kaons which originate from hadronic interactions of cosmic-rays with surrounding matters in cosmic accelerators. 
The detection of high-energy astrophysical neutrinos therefore enables us to address the sources of high-energy cosmic-rays. 
The IceCube Neutrino Observatory is capable of detecting high-energy neutrinos with unique instrumental volume of 1 km$^3$ within the Antarctic ice sheet. 
Recently, blazars as candidate sources of the high-energy neutrino flux are reported by IceCube collaboration~\cite{IceCube:2018cha,IceCube:2018dnn},
and other searches are in progress~\cite{IceCube:2019cia,IceCube:2021xar}.
In addition to the search for sources of high-energy cosmic-rays, optimization for neutrinos with energy in the TeV to PeV range allows IceCube to probe into fundamental physics beyond current laboratory scale~\cite{Lee:1960qv,Glashow:1960zz,Lee:1961jj,Seckel:1997kk,IceCube:2021rpz}. 
In multi-TeV region, inelasticity distributions of neutrino charged-current (CC) scattering are measured using IceCube data of five years~\cite{IceCube:2018pgc}.
The energy-dependence of neutrino-nucleon cross sections above 10 TeV is also extracted from IceCube high energy cascade data~\cite{IceCube:2017roe,Bustamante:2017xuy}.
These measurements admit of the test to Quantum Chromodynamics (QCD) through neutrino interactions at very high energy scales, which interplays with the future FASER$\nu$ experiment~\cite{FASER:2020gpr}.
Besides studies on QCD, new physics searches at high energy scale are performed.
Potential decays of high-energy neutrinos from distant astrophysical sources can significantly alter neutrino flavor composition and this phenomenon can be constrained by measurements of neutrino fluxes~\cite{Beacom:2002vi}. These measurements also allow us to test CPT invariance at high-energy scale~\cite{Hooper:2005jp}. 
More phenomenological researches probing into new physics with neutrino telescopes can be found in~\cite{Lipari:2001ds,Han:2004kq,Beacom:2006tt,Albuquerque:2006am,Yuksel:2007ac,Ng:2014pca,Ioka:2014kca,Shoemaker:2015qul,Murase:2015gea,Bustamante:2016ciw,IceCube:2016dgk,Denton:2018aml,IceCube:2018tkk,ANTARES:2020leh,Bustamante:2020mep}.

\begin{figure}[thbp]
\centering
\includegraphics[width=0.46\textwidth]{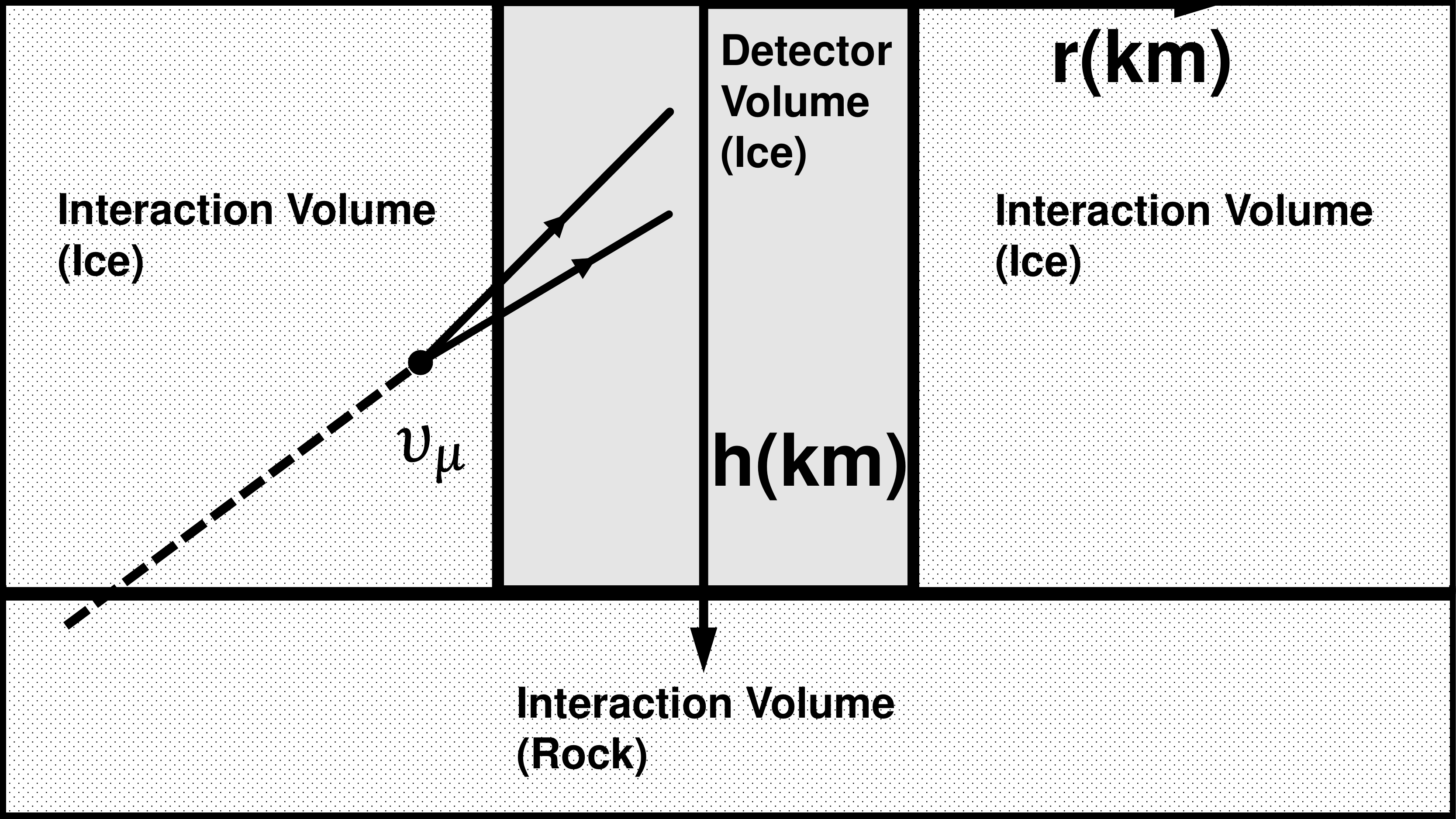}
\caption{Schematic illustration of the simulation framework. The primary muon neutrino interacts outside the detector and produces a pair of muons.  The detector locates above rock media, padded with ice in the horizontal direction.
}\label{fig:detector}
\end{figure}
Apart from the basic single-track topology, widespread interests exist in double-track events~\cite{Bi:2004ys,Kohri:2006cn,Ando:2007ds,Ahlers:2007js,Meade:2009mu,Feng:2010gw,Agashe:2014yua,Kopper:2015rrp,IceCube:2015hle,Feng:2015hja,Ge:2017poy,Fox:2018syq,BhupalDev:2019mon,Zhou:2019frk,Zhou:2019vxt,Meighen-Berger:2020eun,IceCube:2021kod}.
In most beyond standard models (BSM), a $\mathbb{Z}_2$ symmetry is assumed to ensure the stability of the lightest particle, which results in the pair production of new physics particles. 
This scenario has been investigated in supersymmetric models with breaking scale below 10$^{10}$ GeV~\cite{Albuquerque:2003mi}. Below that scale, the typical pair products are charged sleptons. 
Owing to their heavy masses, charged sleptons suffer much smaller energy losses in media, meanwhile their weak coupling to the lightest supersymmetric particle entails them long lifetimes. These two reasons render them a vast production volume, which compensates for their low production rate and results in double-track signals with hundreds of meters in lateral separation. 
Double-track signal with lateral distance of hundreds of meters is extremely rare in standard model (SM). The main background comes from two neutrinos produced in the same air shower, which only contributes to about 0.07 event per year~\cite{vanderDrift:2013zga}.
In SM scenario, the more general source of double-track events generally arises from CC DIS interactions of neutrinos with nucleons. The nucleons are broken apart and produce a charm quark later fragments into charmed hadrons. These hadrons then decay into the second muons
\begin{equation}
	\nu_{\mu} N \rightarrow \mu^- c(D)X \rightarrow \mu^- \mu^+ \nu_{\mu} X.
\label{eqa:dimuon}
\end{equation}
This so-called dimuon process is essential in the determination of strange-quark PDF~\cite{Berger:2016inr,Gao:2017kkx}, and has been measured for decades in various DIS experiments~\cite{osti_879078,NuTeV:2001dfo,NOMAD:2013hbk}.
On the contrary, no signals for such events have been observed in current neutrino telescopes mainly due to their small dimuon lateral distances.
The lateral distance of this process is typically distributed much below 100 meters.
In this paper, a simulation framework down to detector-level is constructed to estimate sensitivities for such dimuon events. To meet the experimental cut for lateral distance, only dimuon productions outside the detector volume are considered.
The dominate background of dimuon events are from the misconstruction of single muon to dimuon. So analysis based on reconstruction algorithms is employed to obtain acceptance rates of dimuons as well as fake rates of single muons. The results indicate that, though fake rates to misconstruction of percent level can be achieved, the large background from single muon events can greatly lower the experimental significance. But this difficulty is surmountable for denser configurations, for instance, KM3NET/ORCA~\cite{KM3Net:2016zxf}.
The recent work~\cite{Zhou:2021xuh} studied a similar scenario without detector-level simulation. We thus also compare the predictions on energy distributions of muons in both works. Furthermore, we discuss the distinctions arsing from different calculations of energy losses.
\begin{figure*}
    \centering
    \includegraphics[width=1.0\linewidth]{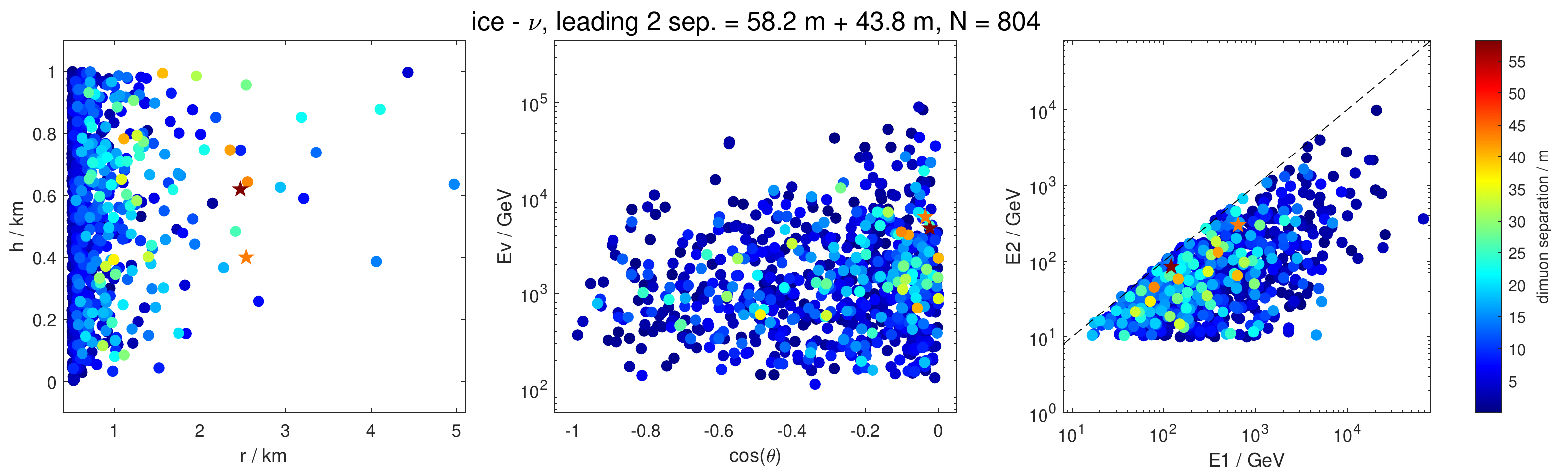} \\
    \caption{
    Statistics for dimuon events induced by atmospheric muon neutrino in ice.
    The exposure time takes ten years.
    Energy cut for primary neutrino is set at 100 GeV, and energy cut for muons at the detector boundary is set at 10 GeV.
    Every colored point represents one dimuon event, and the relevant color stands for the lateral distance of dimuon at the detector boundary.
    Two events with the first and second largest lateral distance are marked with stars.
    {\bf Left:} $h$ and $r$ coordinates of starting positions of dimuon events.
    {\bf Middle:} energy and zenith angle of the primary neutrino that induces the dimuon event. 
    {\bf Right:} energies of dimuon when entering the detector. E1 and E2 represent energies of the leading and sub-leading muon respectively.
    }
    \label{fig:ice_v}
\end{figure*}
\begin{figure*}
    \centering
    \includegraphics[width=0.98\linewidth]{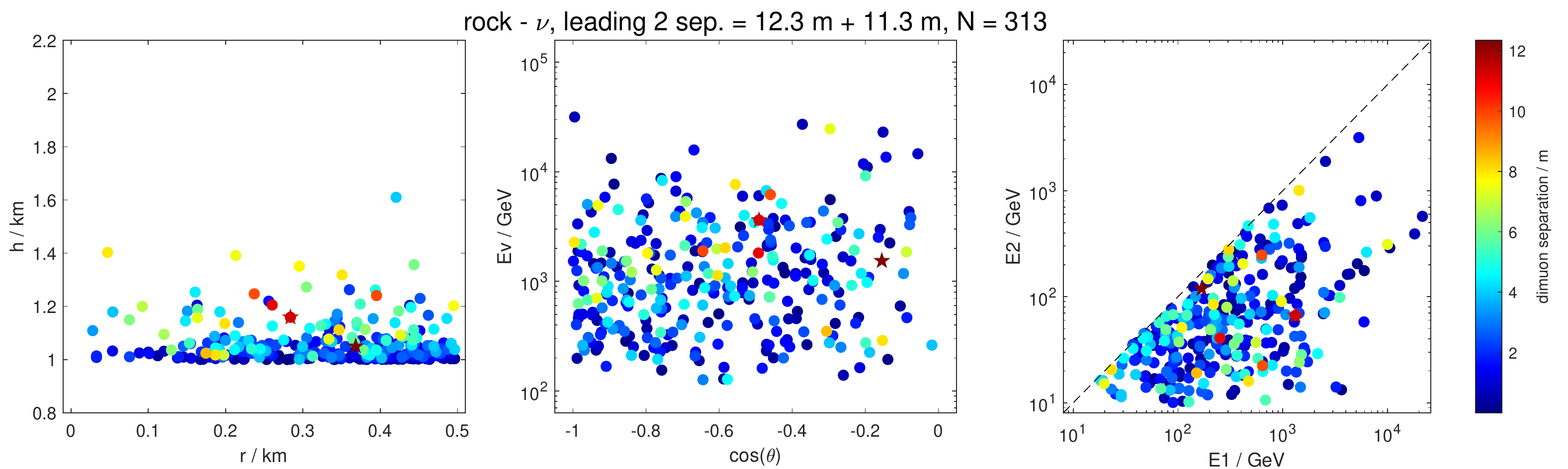} \\
    \includegraphics[width=0.98\textwidth]{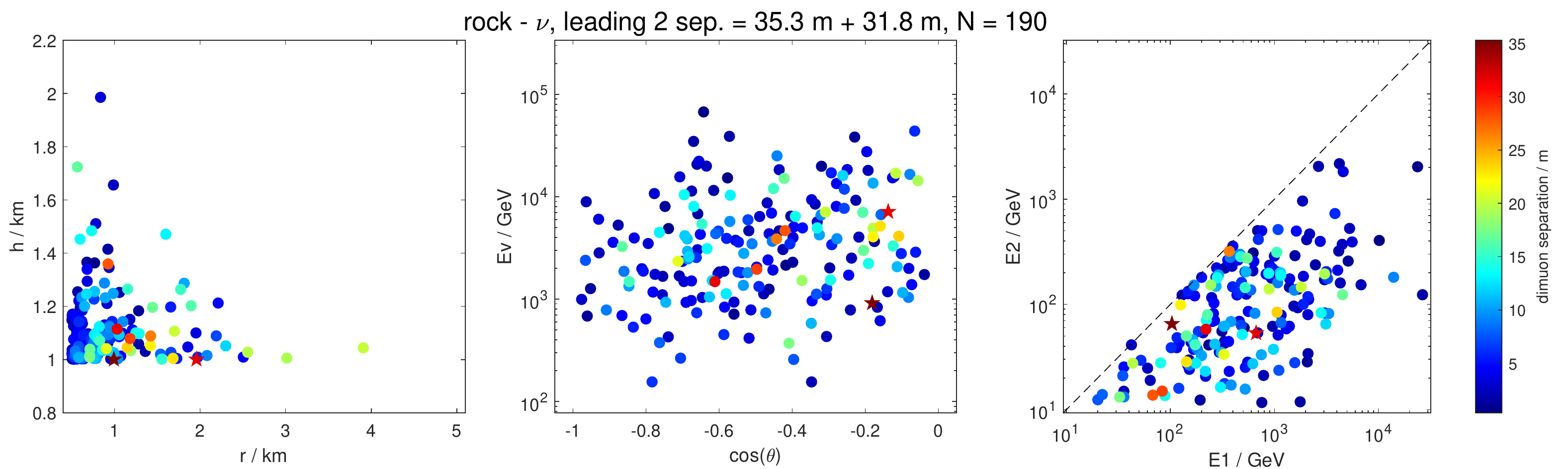} \\
    \caption{The same as Figure~\ref{fig:ice_v}, but for muon neutrino in rock. Due to the different behaviors of energy losses, this interaction volume is divided into two sub-regions with $\rm{r}<0.5\rm{\ km}$ (upper) and $0.5\rm{\ km}<\rm{r}<5\rm{\ km}$ (lower).}
    \label{fig:rock_v}
\end{figure*}
The rest of this paper is organized as follows. 
In Sec.~\ref{sec:simulation}, we discuss the simulation framework.
We elaborate aspects of neutrino flux, and give descriptions of the detector model as well as the event generation method in use.
Simulation results are presented at the end of this section.
We then compare our results with the recent work~\cite{Zhou:2021xuh} in Sec.~\ref{sec:comparison}. Cuts and other conditions are set similar to that work.
In Sec.~\ref{sec:analysis}, the detector simulation is completed. Several detector-level settings are selected to estimate significances of the dimuon signal over the single-muon background.
Finally we conclude with Sec.~\ref{sec:conclusion}.
\section{Simulation Framework}\label{sec:simulation}
The generation of dimuon events follows these steps. At first, event rates are computed in the interaction volume. These rates are taken from the convolution of neutrino flux with relevant cross-sections, multiplied by the exposure time and target number.
Dimuon events are then generated according to differential distributions and normalized to these event rates.
These dimuons are subsequently steered to the detector. If any one in the muon pair loses its whole energy before reaching the detector surface, this event is discarded.
The final sample of dimuon events is the set of dimuons that both can reach the detector boundaries. 
More details are presented as follows.
\subsection{Neutrino Flux}
High-energy neutrino flux can be roughly divided into astrophysical and atmospheric components. 
The former has a harder spectrum than the latter and dominates at high-energy region.
Below 100 TeV, however, atmospheric flux is three to four orders larger than the astrophysical flux~\cite{IceCube:2013gge}.
Given the approximate linear relation of neutrino-nucleon cross-section to neutrino energy in the range of tens of GeV to a few TeV , spectral indices 3.7 and 2 for atmospheric and astrophysical flux, respectively, 
the DIS event rate of atmospheric neutrinos in this energy range is three orders of magnitude larger than their astrophysical counterpart.
This estimation matches with simulation of DIS muon events in~\cite{IceCube:2003llu}. 
Consequently, as a subset of DIS muon events, dimuon events from atmospheric sources are also expected to have higher order of magnitude.
For energy above 100 TeV, astrophysical flux becomes dominant because of the softer decreasing behavior in event number of $E^{-1}$ compared with $E^{-2.7}$ for atmospheric flux. 
But the absolute strength of flux is negligible in this region, and the neutrino-nucleon cross-section practically drops down from the linear increment~\cite{Cooper-Sarkar:2011jtt,Bertone:2018dse,Gao:2021fle}.
The suppression becomes stronger when the neutrino absorption resulting from the propagation through Earth is taken into account. Such phenomenon becomes sizable for neutrinos with energies higher than several hundreds TeV~\cite{Albuquerque:2001jh}.
We therefore keep only neutrino flux from atmospheric component in this work. A posterior estimation shows that dimuon events are mainly induced by neutrinos with energies below 10 TeV, in which flux attenuation due to the Earth absorption can also be safely neglected.
On the other hand, neutrino oscillation leads to the disappearance of muon neutrinos. This effect, however, only impacts on neutrinos with energies below 100 GeV~\cite{IceCube:2017zcu}, and the energy cut allows us to exclude this region from our analysis.
The atmospheric neutrino flux below 10 TeV is taken from HKKM15~\cite{Honda:2015fha},
and is extrapolated up to 1 PeV by fitting a standard parameterization in~\cite{IceCube:2013gge}.
To reduce atmospheric muon background, only up-going neutrinos (zenith angle larger than 90$^\circ$) are kept.  
Specially, the detailed treatment of very high energy neutrino induced dimuon events would be more complicated. In such scenario, contribution from astrophysical neutrinos, prompt atmospheric neutrinos as well as flux attenuation due to absorption of Earth play essential roles.
For ultra-high energy, subtle treatments of calculation of cross-sections are also necessary~\cite{Bertone:2018dse}.
\subsection{Detector and Event Rates}
The detector is modeled as a cylinder with height of 1 km and radius of 0.5 km. 
For the convenience of description, we set a cylindrical coordinate displayed in Figure~\ref{fig:detector}, in which the origin is at the center of the top surface of detector and the unit of length is given in kilometer.
The detector volume then covers $\{r < 0.5, 0 < h < 1\}$ and consists of ice.
Because the interaction volume is outside the detector volume, it is either in region $\{r < 5, 1 < h < 2\}$ filled with rock, or $\{0.5 < r < 5, 0 < h < 1\}$ filled with ice. The densities of ice and rock are taken to be 0.92 $\rm{g}/\rm{cm}^3$ and 2.70 $\rm{g}/\rm{cm}^3$, respectively. The chemical composition of ice is $\rm{H_2O}$, and rock is composed of elements same as the Earth crust.
%

%
The first step is to calculate the total event rates from
\begin{equation}
\frac{d^2N_{\rm{tot}}}{dE_\nu d\theta}
    = N T \times
    2\pi \frac{d^2F_\nu}{dE_\nu d\theta}
    \times \sigma_{\rm{tot}}(E_\nu)
\label{eqa:event_rates}
\end{equation}
where $N$ is the number of nucleons in the interaction volume, $T$ is the exposure time, and $d^2F_\nu/dE_\nu d\theta$ is the differential neutrino flux over neutrino energy and zenith angle.
$\sigma_{\rm{tot}}$ is the total cross-section per nucleon for neutrino CC DIS taken from the integration of
\begin{equation}
\begin{aligned}
\frac{d^2 \sigma ^{\nu(\bar{\nu})} _{\rm{cc}}}{d \textit{Q}^2 d\textit{W}^2}
    &= \frac{G_F^2 m_W^4}{2 \pi} \frac{1}{(\textit{Q}^2+\textit{W}^2)(\textit{Q}^2+m_W^2)^2} \\
    & \ \ \ \bigg( (1-\textit{y})F_2(\textit{x},\textit{Q}^2) + \textit{y}^2 \textit{x}F_1(\textit{x}, \textit{Q}^2) \\
    & \ \ \        \pm \textit{y}(1-\frac{\textit{y}}{2}) \textit{x}F_3(\textit{x}, \textit{Q}^2) \bigg),
\label{eqa:ccdis}
\end{aligned}
\end{equation}
for either charm-quark production or inclusive production in hadronic final states,
and plus (minus) sign accounts for $\nu$ ($\bar{\nu}$). Q$^2$ is the transferred momentum square and W is the invariant mass of the final hadronic system. x and y denote the Bjorken-x and inelasticity of the DIS process, respectively. 
In our calculation, the structure function $F_i(x, Q^2)$ is adopted at next-to-leading order in QCD~\cite{Gao:2017kkx} to reduce the perturbative uncertainties.
Additionally, distinctions between ice and rock target are taken into account and CT14NNLO PDF set~\cite{Dulat:2015mca} is used.
Neutrino scattering events are generated according to the aforementioned differential distribution. Their production positions are distributed uniformly since the mean free path of neutrino is much larger than the length scale of the interaction volume, and the azimuthal angles of incident neutrinos are also distributed evenly by the definition of the flux model.
These DIS events are then fed into {\tt Pythia v8}~\cite{Sjostrand:2014zea} for parton showering, hadronization and hadron decays,
from which dimuon events are generated and their energies along with separation angles are extracted.
The primary muon comes from the DIS primary vertex, and is, in most of the time,  more energetic,
whereas the second muon is the decay product of charmed hadrons.
The displacement between two production vertices is much smaller than the lateral distance of dimuons at the detector boundary and is, therefore, neglected. Hence, we take the starting position of dimuons as the neutrino interaction point. 
Another approximation is in the direction of two muons.
Due to large boost, dimuons follow the direction of primary neutrinos to a good approximation. Their separation angles $\theta_{\nu\mu}$ obey a relation $\tan\theta_{\nu\mu} \approx \langle P_\perp \rangle / E_\nu$, where $\langle P_\perp \rangle$ is the average transverse momentum of the primary muon~\cite{FASER:2019dxq}.
For $E_\nu = 1 \text{ TeV}$, we have $\langle P_\perp \rangle \sim 30 \text{ GeV}$, and $\theta_{\nu\mu}$ reaches a value $0.03 \ll 1$.
This is also true for the separation angle $\theta_{\mu\mu}$ of two muons. For $E_\nu \gg \text{a few GeV}$, $\theta_{\mu\mu}$ is of order $p_{CM}/p_{LAB} \approx \sqrt{\text{GeV}/E_\nu} \ll 1$. The lateral distance of dimuon is then simplified as $D_{\mu\mu} \approx \theta_{\mu\mu} l$, where $l$ is the distance dimuon traversed.
At last, dimuon is steered to the boundary of the detector. The energy loss is simulated with {\tt MMC v1.6.0}~\cite{Chirkin:2004hz}.
These dimuon events that either of two muons cannot reach the detector boundary are discarded. The dimuon lateral distance at the boundary, denoted as dimuon separation, as well as other kinematic information of two muons are recorded.
It is noted that there is not much ambiguity from the choice of interaction volume.
In fact, the interaction volume should be taken large enough to encapsulate detectable dimuon events as many as possible.
For an ideal and exact simulation, the interaction volume can be taken to cover the whole Earth.
The distribution of production positions in this case is then impossibly uniform. But this simulation will be very time-consuming and very large part of the interaction volume is useless because events generated there cannot reach the detectable region.
It is therefore useful to adopt the concept of flux attenuation responsible for events that are clearly impossible to reach the detector.
In our consideration, the view of attenuation continues until the flux reaches the interaction volume, and the losses from this attenuation are small and ignored.
On the other hand, the interaction volume must not be taken so small as to lose potential dimuon events that can reach the detector.
It can be seen from Figure~\ref{fig:ice_v} and Figure~\ref{fig:rock_v} that only a small portion of survived events are generated at the far-end of the interaction volume.
This convinces us that our choice of the interaction volume is indeed large enough to include most if not all detectable dimuon events.
\subsection{Simulation Results}
One sample in our simulation framework for muon neutrino is demonstrated in Figure~\ref{fig:ice_v} and Figure~\ref{fig:rock_v}. Results for muon anti-neutrino are similar and not shown here.
The exposure time is ten years.
Cuts on primary neutrino energies and muon entering-detector energies are set to be 100 GeV and 10 GeV, respectively.
Statistics for dimuon events induced by muon neutrinos in the interaction volume filled with ice are shown in Figure~\ref{fig:ice_v}.
Each colored dot represents one dimuon event, and the kinematic information is given by relevant coordinates.
The dot color denotes the lateral distance at the time of dimuon entering the detector, or say, the dimuon separation.
Most dimuon events have separations below 30 meters, and two largest separation are marked with stars.
A total number of 804 events are collected, with two largest dimuon separation 58.4 m and 43.8 m.
The left panel shows the starting positions of dimuon events. Almost all events occur near the detector and in region $0.5 < r < 2.5$.
This is because lower-energy muons at starting positions are mostly distributed below several hundred GeV and on average a 500 GeV muon exhausts its energy in ice when travelling about 2 km. 
The right panel of Figure~\ref{fig:ice_v} shows energy distributions of two muons when they enter the detector. 
E1 and E2 represent energies of the leading and sub-leading muons respectively.
In spite of energy losses before reaching the detector, a significant amount of sub-leading muons have energy smaller than a few hundred GeV.
In the middle panel, the energies and zenith angles of the primary neutrinos inducing the dimuon events are shown.
The preference of $\cos(\theta) = 0$ is a reflection of the fact that neutrino flux peaks at the horizon.
The same information for interactions in rock is displayed in Figure~\ref{fig:rock_v}.
Because muons can lose energies both in the ice and in the rock for $0.5<\rm{r}<5$, while they only lose energies in rock for $\rm{r}<0.5$, we divide interaction volume of rock into two sub-regions, and the statistics are shown in the lower and the upper panels, respectively. For $\rm{r}<0.5$ in the upper panel, events have rather small separations and prefer directions close to up-warding. Contributions from this region are, then, negligibly small. For $0.5<\rm{r}<5$ in the lower panel, some events reach the detector with slightly larger dimuon separations. However, the total number of events turn out to be small. So the interaction volume of rock only gives sub-leading contributions.
\section{Comparison}\label{sec:comparison}
Both the recent work~\cite{Zhou:2021xuh} (labelled as ZB21) and this paper discuss a similar situation.
It is therefore worth comparing results from both works.
In this section, comparison of dimuon events for ten years exposure of IceCube is performed.
We also discuss the difference between two calculation methods of muon energy losses.
\subsection{Comparison of 10 Years Through-going Dimuons}
\begin{figure}[t]
    \includegraphics[width=0.50\textwidth]{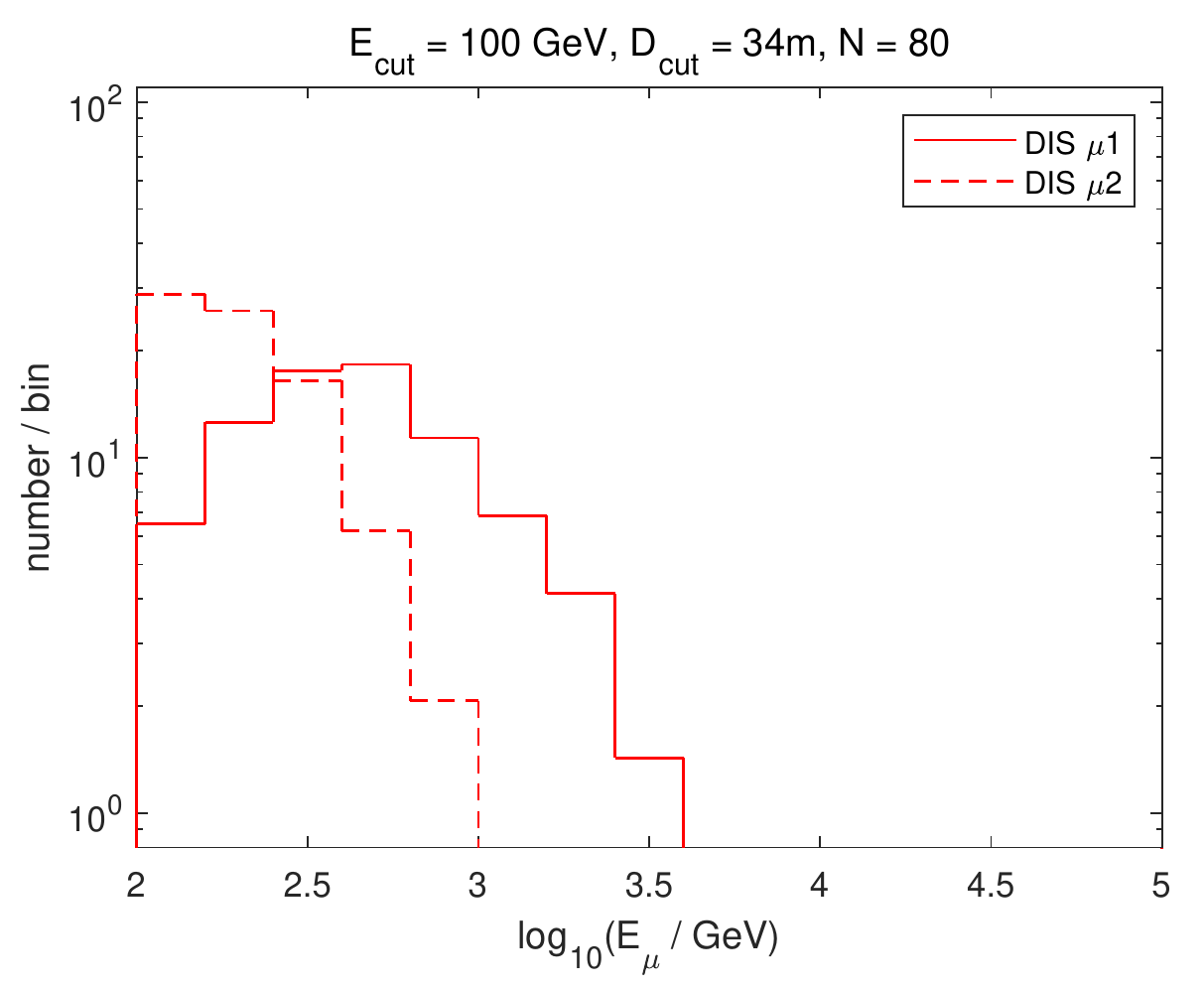}
    \caption{
    Predictions of energy distribution of two muons when entering the detector.
    Exposure time is ten years. $\mu$1 is defined to be the more energetic muon.
    The same cut as Eq.~(7) in~\cite{Zhou:2021xuh} is set.
    After considering the difference between cross-sectional area, this figure can be directly compared with the right panel of Figure~5 in~\cite{Zhou:2021xuh}. 
    }
    \label{fig:comp_10y}
\end{figure}
Except for the simulation framework, distinctions from some setups between ZB21 and this paper are listed as follows:
$\bullet$ Zenith angle in ZB21 covers the whole sky, while only up-going neutrinos are considered in this paper.
$\bullet$ In ZB21 a cross-sectional area of 1 km$^2$ is used in the analytical calculation, which has no direct
correspondence in our simulation framework. We estimate an effective area of 0.85 km$^2$ that is calculated from the two-thirds power of
our detector volume.
$\bullet$ Average energy loss formula is used in ZB21, but in this paper, muon energy loss is simulated with Monte Carlo (MC) program {\tt MMC}.
$\bullet$ In ZB21, atmospheric neutrino flux above 10 TeV and astrophysical neutrino flux are both taken from IceCube's measurement. But we only take into account atmospheric part and use standard parameterization to extrapolate it. Flux attenuation due to Earth absorption is treated in ZB21 but not in this work.

\begin{figure}[!t]
    \includegraphics[width=0.5\textwidth]{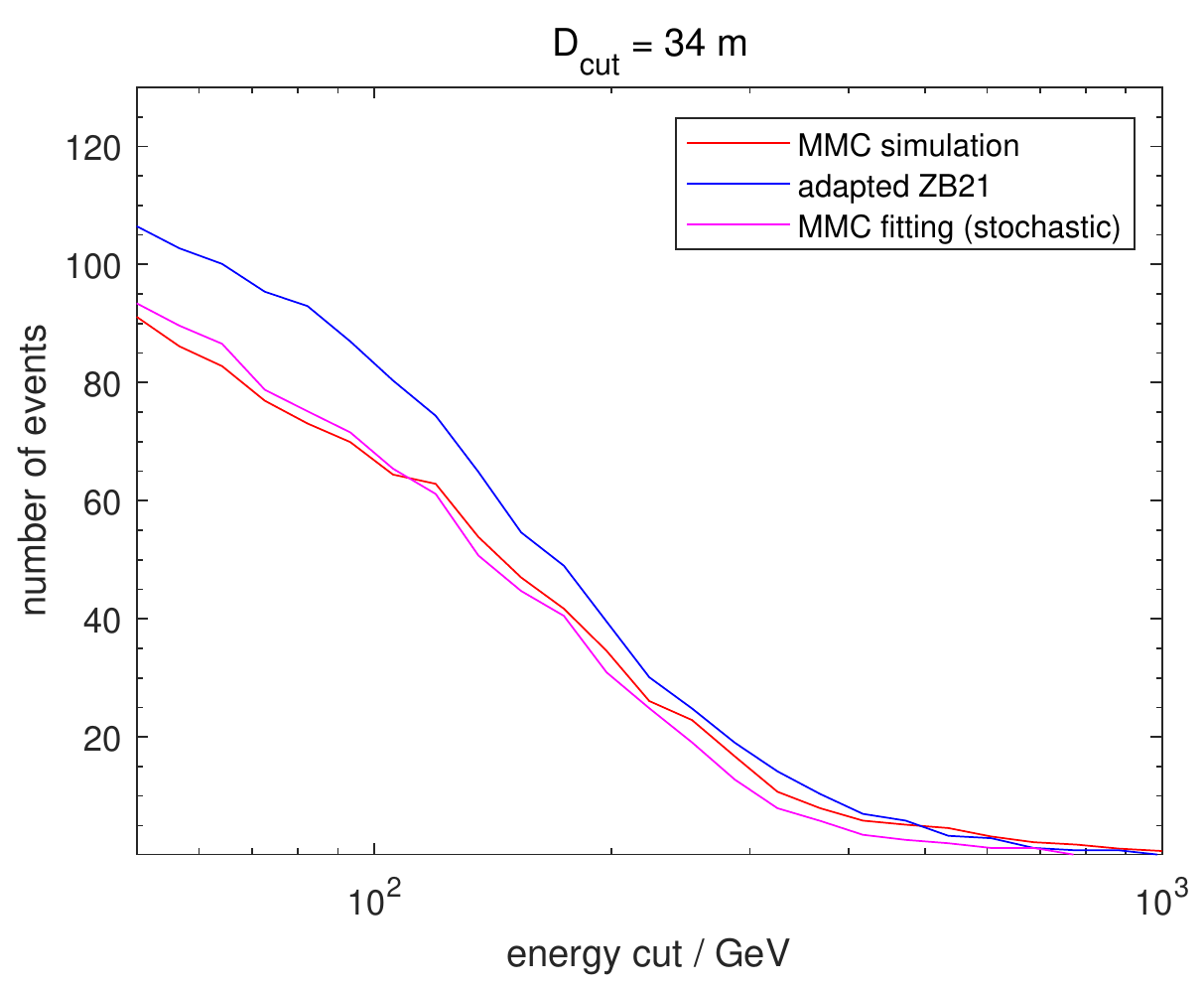}
    \caption{
    Comparison of number of events passing cuts for two different approaches.
    MMC simulation (average energy loss formula using parameters in~\cite{Zhou:2021xuh}) is used to calculate the energy loss for red (blue) curve.
    Curve using average energy loss formula based on parameters by {\tt MMC} stochastic fitting is also plot in magenta.
    x axis is energy cut for muon at the detector boundary, and y axis indicates the number of events passing the cut as well as the lateral distance cut which is set to 34 meters.
    }
    \label{fig:zb21e_vs_oure}
\end{figure}
The first distinction is eliminated by doubling the events in each bin since an approximate reflection symmetry exists.
We do not rescale our binned events number by $1/0.85$ to eliminate the second one. But this can be done if only the total event number matters.
Influence on normalization from the last distinction is expected to be small since both works predict a small contribution from neutrino energy higher than 10 TeV.
But differences in differential energy distribution at the high energy tail can be marked, especially that for the leading muon.
We completely remove the third difference by using the average energy loss formula $dE/dX=-\alpha - \beta E$ since the starting position of dimuon in this comparison.
Energies of two muons at the detector boundary are then calculated based on that formula, as well as the distance muon can traverse.
The comparison is performed for ten years exposure of IceCube. Our result is shown in Figure~\ref{fig:comp_10y}, and the counterpart is the right panel of Figure 5 in ZB21.
The lateral distance cut is set to be the same as ZB21's, i.e., $R_{\mu2}\theta_{\mu\mu} > 34 \text{ m}$. Here, $\theta_{\mu\mu}$ is the separation angle of dimuon, and $R_{\mu2}$ is the range of the sub-leading muon between the dimuon starting position and where it exits the detector or stops in it.
The cut on muon energy when entering the detector is also set to be the same as ZB21's, namely 100 GeV.
The total number of dimuon events passing the cut is 80. After rescaling by ratio of cross-sectional area $1/0.85$, we obtain a value 94, which is about 110\% of that in ZB21.
However, such a rescaling is rather informative than exact due to the ambiguity in definition of
the effective area on the detector geometries.
Energy distributions for leading and sub-leading muons follow similar patterns in both works,
whereas differences can be seen for muon energy greater than about 1 TeV.
Apart from the fluctuation of Monte Carlo computation, this difference may come from different treatments in high energy flux.
One noteworthy feature is that lateral distances of dimuons can be markedly amplified inside the detector,
which manifests in the fact that $R_{\mu2}\theta_{\mu\mu}$ is larger than dimuon separation at the detector boundary by a factor of 2 to 5 or more.
This implies the approximation of parallel dimuon is crude at the level of tens of meters,
and it is therefore conservative to use dimuon separation at the detector boundary as a criterion for experimental lateral distance cut in dimuon searches.
\subsection{Distinctions from Energy Losses Calculation}
Besides the complete removal of the third difference by the employment of average energy loss formula, we also count dimuon events passing the energy cut at the detector boundary with {\tt MMC} simulation.
And the lateral distance cut is still conducted the same way as ZB21.
The results are illustrated in Figure~\ref{fig:zb21e_vs_oure}.
In that figure, x-axis indicates the muon energy cut at the entering-detector point, and y-axis is the number of dimuon events passing both the energy and the lateral distance cuts.
The blue curve (adapted ZB21) is for the method used in the previous subsection.
We also consider whether dimuon can pass the energy cut with {\tt MMC} simulation, which is shown in red curve.
Note that {\tt MMC} simulation is only used for predicating whether dimuon can pass the energy cut, final energy losses that decide the lateral distance are still consistently calculated with average energy loss formula with the same parameters as the previous subsection.
From this comparison, it is observed that adapted ZB21 leads to enhancement of dimuon events in comparison with the {\tt MMC} simulation.
For low energy thresholds, the relative decrements are 10\% at most, and the discrepancy in absolute number becomes smaller as the energy cut increases. 
One reason for this behavior is the fitted parameters $\alpha$ and $\beta$ in {\tt MMC} are different from that taken in ZB21.
In their work, $\alpha$ and $\beta$ are taken to be 3.0 $\times$ 10$^{-3}$ GeV cm$^2$/g and 3.0 $\times$ 10$^{-6}$ cm$^2$/g, respectively\footnote{We note that the first two versions of ZB21 on arXiv used parameters $\alpha$ of 2.0 $\times$ 10$^{-3}$ GeV cm$^2$/g and $\beta$ of 3.0 $\times$ 10$^{-6}$ cm$^2$/g. For low energy thresholds, this choice leads to at most 40\% relative decrements of events passing cuts compared to {\tt MMC}.}.
Furthermore, the stochastic process makes muons travel shorter distance in general~\cite{Chirkin:2004hz}.
To explain this point more clearly, similar curve based on $\alpha$ and $\beta$ from {\tt MMC} stochastic fitting is plot in magenta,
in which $\alpha$ = 2.68 $\times$ 10$^{-3}$ GeV cm$^2$/g and $\beta$ = 4.70 $\times$ 10$^{-6}$ cm$^2$/g.
This curve is very close to that from Monte Carlo simulations.
Uncertainties from the modeling of processes like pair production, photonuclear interactions and bremsstrahlung in {\tt MMC} are small~\cite{Chirkin:2004hz}.
\section{Experimental Side}\label{sec:analysis}
In this section, a phenomenological study on the experimental potential probing dimuon events is performed.
A framework of simplified detector response modelling is constructed. Based on it, multiple simulation settings are applied to estimate dimuon acceptance efficiencies.
Considering that a large fraction of background are mis-reconstructed single-muon events, single-muon fake rates are also calculated in corresponding situation.
Multiplied by the detector-level evaluation, significances are estimated for several theoretical benchmarks.
We simply assume the detector-level settings and theoretical benchmarks can be factorized in the calculation of significance.
In addition to an IceCube-like detector configuration, an ORCA-like~\cite{KM3Net:2016zxf}, as the prototype of denser but smaller detector, is modeled as well.
The estimated results imply that, ORCA shows a larger potential to observe dimuon signals with low energies and short lateral distances than IceCube.

{\renewcommand{\arraystretch}{1.3} 
\begin{table}[b]
\begin{tabular}{|c|c|c|c|c|c|}
\hline
\multicolumn{6}{|c|}{Sparse Configuration (IceCube-like)} \\
\hline
\multicolumn{3}{|c|}{\textit{Set 1}} & \multicolumn{3}{c|}{\textit{Set 2}}\\
\hline
sep. cut & 2*200 $\gev$ & 400 $\gev$ & sep. cut & 2*400 $\gev$ & 800 $\gev$ \\
\hline
18 m & $19.0\%$ & $2.9\%$ & 17 m & $45.5\%$ & $2.5\%$ \\
\hline \hline
\multicolumn{6}{|c|}{Dense Configuration (ORCA-like)} \\
\hline
\multicolumn{3}{|c|}{\textit{Set 3}} & \multicolumn{3}{c|}{\textit{Set 4}}\\
\hline
sep. cut & 2*30 $\gev$ & 60 $\gev$ & sep. cut & 2*50 $\gev$ & 100 $\gev$ \\
\hline
4 m & $57.12\%$ & $0.74\%$ & 4 m & $69.32\%$ & $0.78\%$ \\
\hline \hline

\end{tabular}
\caption{
The optimal dimuon acceptances $\epsilon_{\mu\mu}$ and single-muon fake rates $\epsilon_{\mu}$ with corresponding separation cuts are listed for two type detector configurations. In each configuration, two sets of muon samples with same fixed track separation ($30\,\text{m}$ for sparse, $10\,\text{m}$ for dense) but different energies are reconstructed. In each set, dimuon sample is indicated by a prefactor 2.}
\label{tab:efficiencies}
\end{table}}

\subsection{Detector-level Simulation}\label{sec:eff}
To investigate the effect of detector response on the dimuon identification, the detector simulation and reconstruction of dimuon events are performed. The simulation set up used in this section is modified based on the development in \cite{Hu:2021jjt}. The benchmark detector configuration is composed of 5000 DOMs arranged on 100 strings with 100 meter horizontal and 20 meter vertical spacings, respectively. Each DOM is a 17-inch glass sphere housing 31 3-inch PMTs. Note that, such mDOM design differs from the one in IceCube. However, to perform a fast detector response modelling, this distinction is inevitable. In this simulation, parallel muons are injected horizontally to the detector with fixed separation (30$\,\text{m}$) but different energies and propagated in sea water instead of ice as for IceCube. Simulation settings of two sets are described below: 

$\bullet$ \textit{Set 1}: Two muons in dimuon event have identical energies of 200 $\gev$ with fixed separation $d=30\,\text{m}$, while the single muon background is 400 $\gev$. 

$\bullet$ \textit{Set 2}: Two muons in dimuon event have identical energy of 400 $\gev$ with fixed separation $d=30\,\text{m}$, while the single muon background is 800 $\gev$. 

The single-muon event reconstruction chain starts with a simple line-fit and m-estimator to obtain a better seed for further reconstruction and selection. The m-estimator result is then applied in the first Single-Photon-Electron (SPE) maximum-likelihood reconstruction that only uses the first photon hit on each DOM. The first photon SPE fit is then, followed by a more comprehensive Multiple-Photon-Electron (MPE) maximum-likelihood reconstruction.

\begin{figure}[t]
    \includegraphics[width=0.46\textwidth]{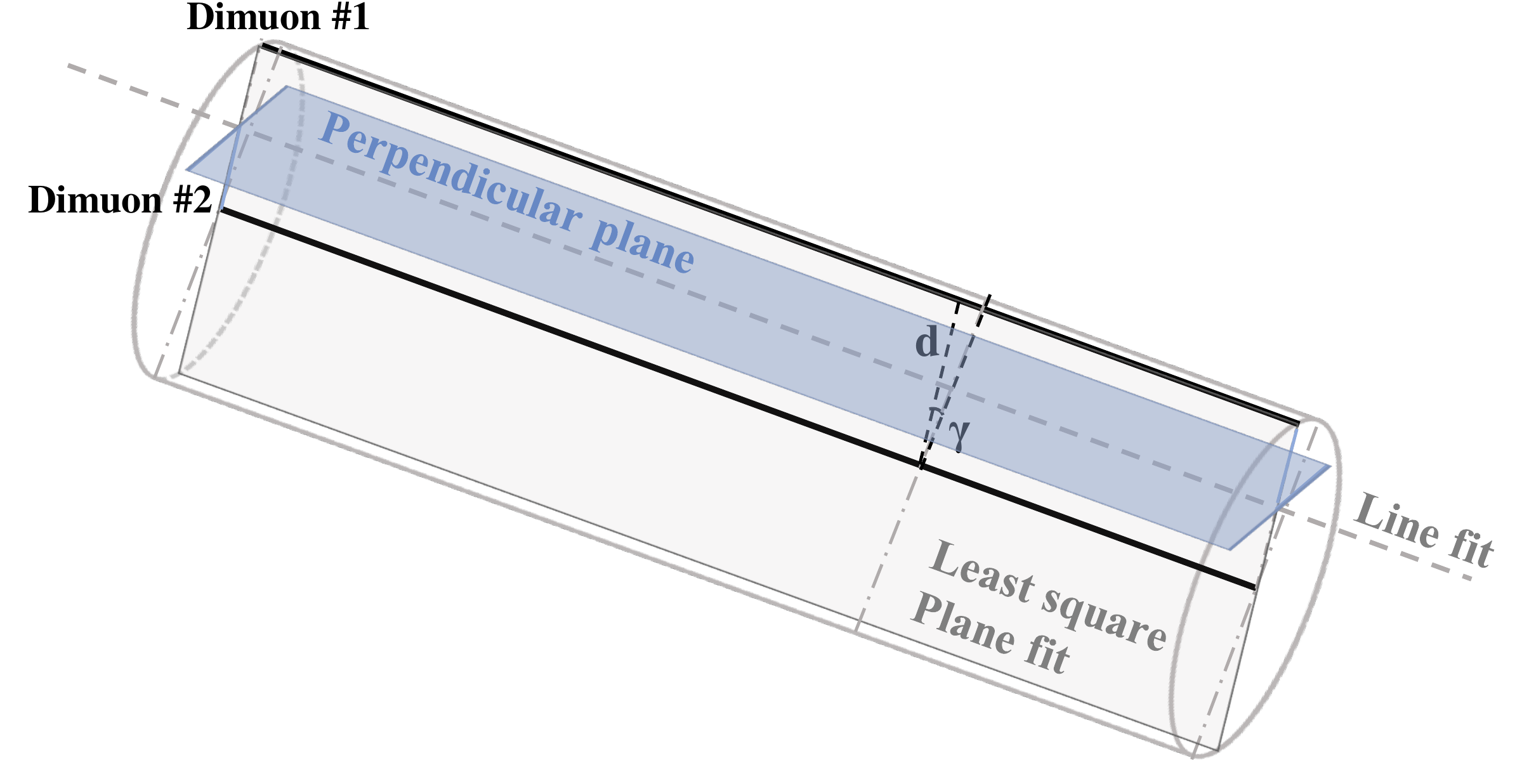}
    \caption{Schematic illustration of dimuon hits splitting. Two parallel muon tracks, indicated as black solid lines, propagate horizontally in the detector. Two additional reconstructed parameters of dimuon hypothesis, track separation $d$ and displacement angle $\gamma$ are marked in the Figure. All hits are splitted into two groups by the perpendicular plane (blue shaded).}
    \label{fig:plane}
\end{figure}

The dimuon reconstruction algorithm is essentially the same as the single-muon reconstruction, but has two additional parameters. The single-muon reconstruction is based on 5 parameters. Three of them are track starting position $(x, y, z)$ and two are track direction $(\theta, \phi)$. In addition to aforementioned 5 parameters, dimuon events have two extra parameters, the track separation $d$ and the displacement angle $\gamma$ of the second track. To construct the dimuon reconstruction algorithm, hits splitting into two groups is necessary. A schematic illustration is displayed in Figure~\ref{fig:plane}. The hits splitting starts with a simple least square plane fit that is similar to the line-fit. Then, a perpendicular plane that contains the line-fit result is constructed and thereby, all photon hits are divided into two sides of this perpendicular plane. 

\begin{figure}[!t]
    \includegraphics[width=0.48\textwidth]{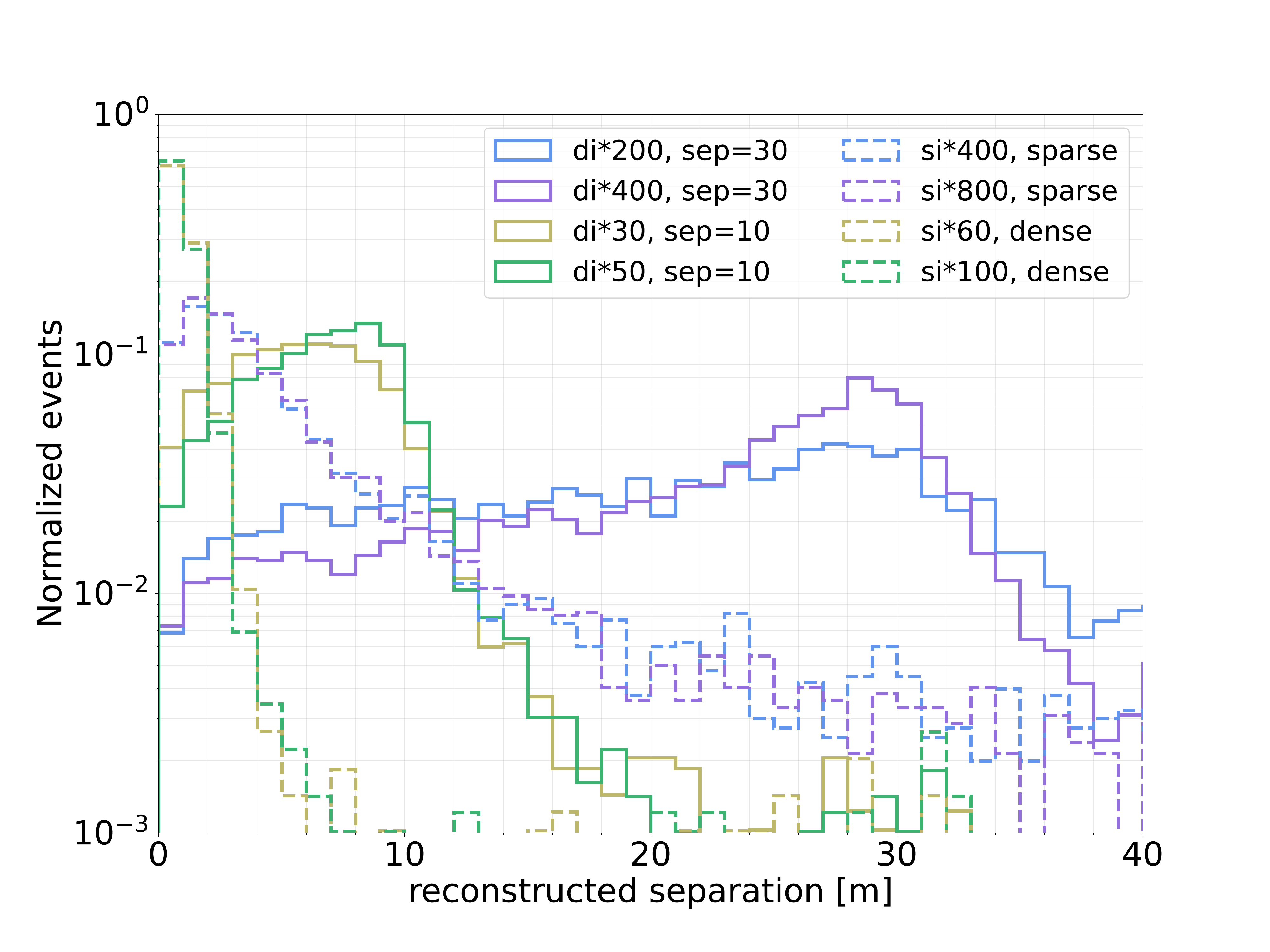}
    \caption{Distributions of reconstructed separations of true single- and di-muon samples, shown by dashed and solid lines, respectively. Samples with 30-meter separation are reconstructed in sparse detector configuration. Samples with 10-meter separation are reconstructed in dense detector configuration. In each set labelled with same color, \textit{di} denotes dimuon and \textit{si} denotes single muons.}
    \label{fig:reco_sep_dis}
\end{figure}

All Monte Carlo events comprising of true single- and di-muon samples are reconstructed with dimuon hypothesis. Depending on the basic reconstruction quality and results, the \textit{pre-level} selection requires that the reconstruction fit converges. The reconstructed separation distributions of true single- and di-muon samples are shown in Figure \ref{fig:reco_sep_dis},
in which blue and purple lines represent \textit{Set 1} and \textit{Set 2}, respectively.
It can be clearly seen that a large fraction of single muon events are misclassified as dimuons with reconstructed separation below about 10 meters. As the separation increases, the dimuon events dominate the distribution.
Therefore, events are further selected by their reconstructed separations and this is referred to as \textit{level-1}. We define a test statistics for events that pass the \textit{level-1} selection: $TS:= -2\ln{(L_{d=0}/L_{bf})}$, where $L_{d=0}$ indicates the likelihood when the separation $d$ is fixed to be zero and $L_{bf}$ indicates the best fit dimuon hypothesis. The event is classified as a dimuon event if its $TS>10$, otherwise the event goes into single-muon sample (\textit{level-2}).

The dimuon acceptance rate is defined as the number of dimuon events that pass the final selection divided by the total injected MC events, while mis-identified single muons are background events.
The reconstructed separation cut is selected to optimize dimuon acceptance rate $\epsilon_{\mu\mu}$ and single muon fake rate $\epsilon_{\mu}$.
Resultant optimized values are listed in Table \ref{tab:efficiencies} with corresponding separation cuts.
As a comparison, a denser but smaller detector configuration similar to KM3NeT/ORCA is additionally investigated, as the medium and DOM design used in this work is the same as ORCA's. In this dense configuration, 20 and 10 DOMs are separated by 10 and 20 meters vertically and horizontally, respectively. In addition to detector geometry, energies of true dimuon samples are reduced to $30\,\gev$ and $50\,\gev$ with 10-meter true separation, whereas the corresponding single muon backgrounds are $60\,\gev$ and $100\,\gev$. They are referred to as \textit{Set 3} and \textit{Set 4}.
Reconstructed separation distributions for these two sets are shown in yellow and green lines in Figure \ref{fig:reco_sep_dis}, and optimized dimuon acceptance rate and single muon fake rate are also listed in Table~\ref{tab:efficiencies}.
By contrast with the sparse detector, the dense detector shows a significant improvement in dimuon classification in spite of reduced track separation, along with reduced single muon fake rates.
In addition to dimuon samples listed in Table \ref{tab:efficiencies}, two muons with unequal energies are explored as well. However, the acceptance rates of dimuons and fakes rates of single muons in this case tend to be closer. This is because, so far, the energy reconstruction is not included in the algorithm. To enhance the track separation, the energy reconstruction, along with the optimization of hits splitting will be taken into account in near future. Moreover, the optimized design of optical sensor, for instance, a hybrid DOM consisting of not only PMTs, but also silicon photomultipliers (SiPMs) with better timing performance would potentially improve the track separation resolution.

\begin{figure*}[t!]
    \includegraphics[width=0.49\textwidth]{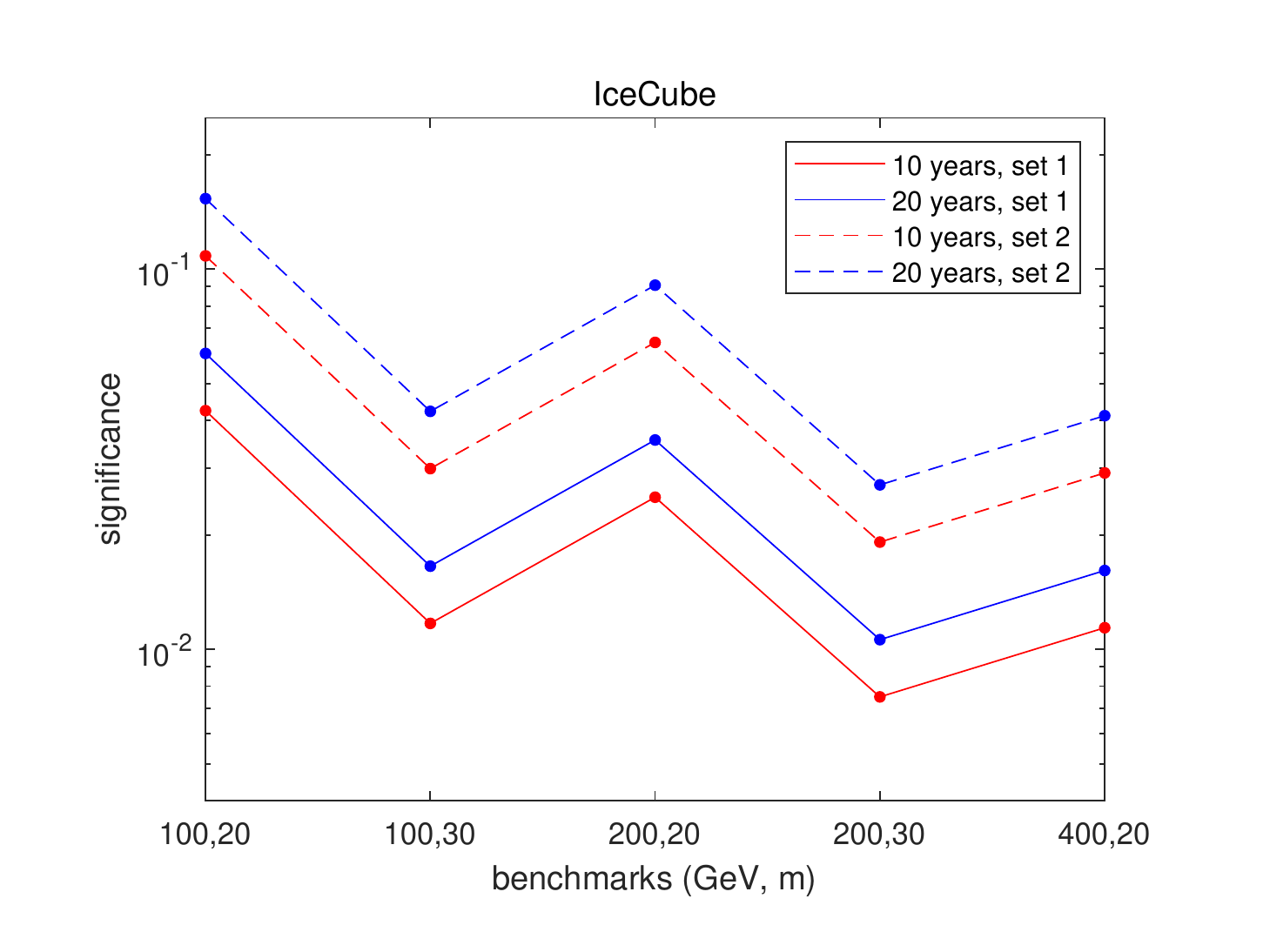}
    \includegraphics[width=0.49\textwidth]{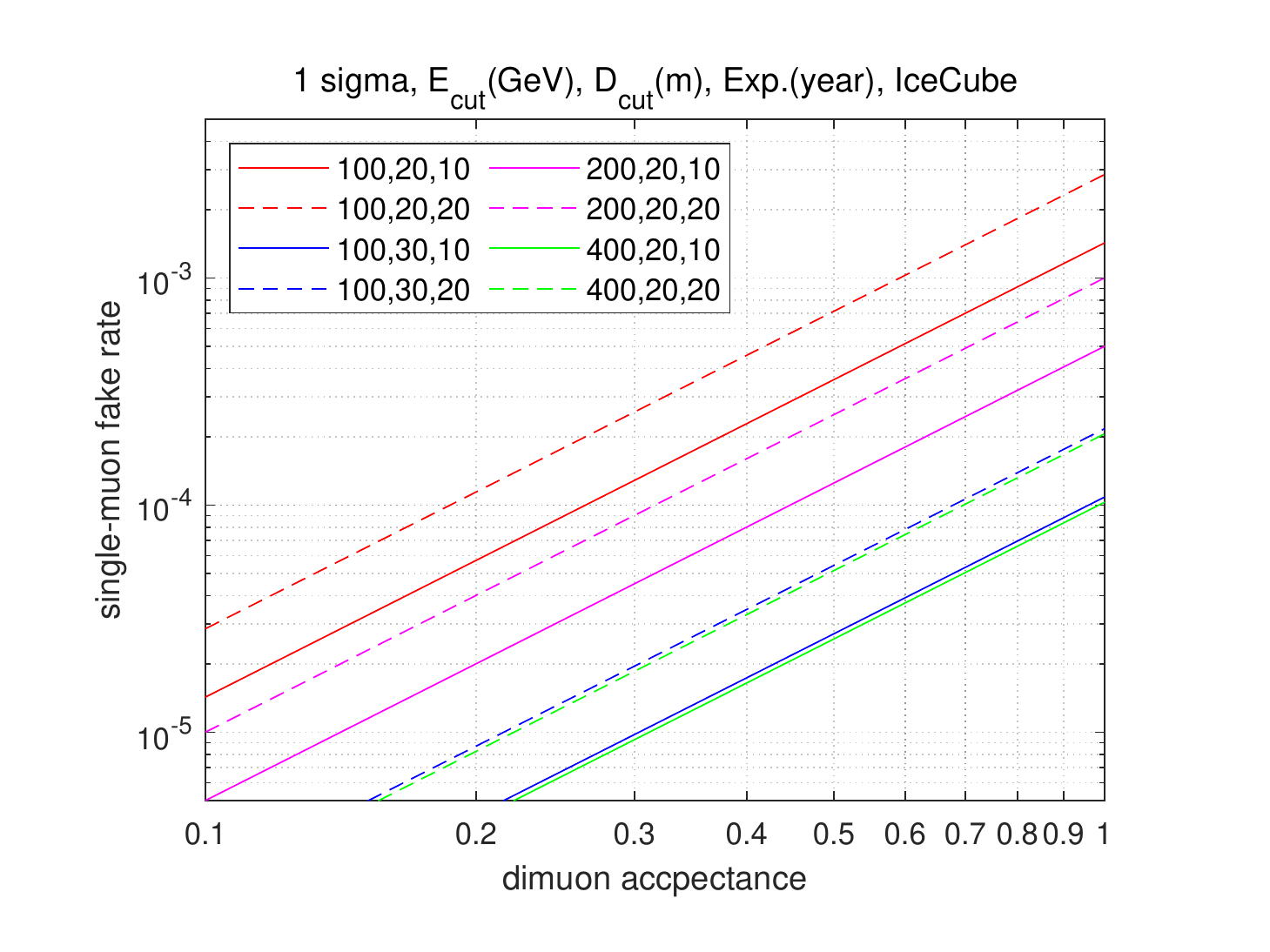}
    \includegraphics[width=0.49\textwidth]{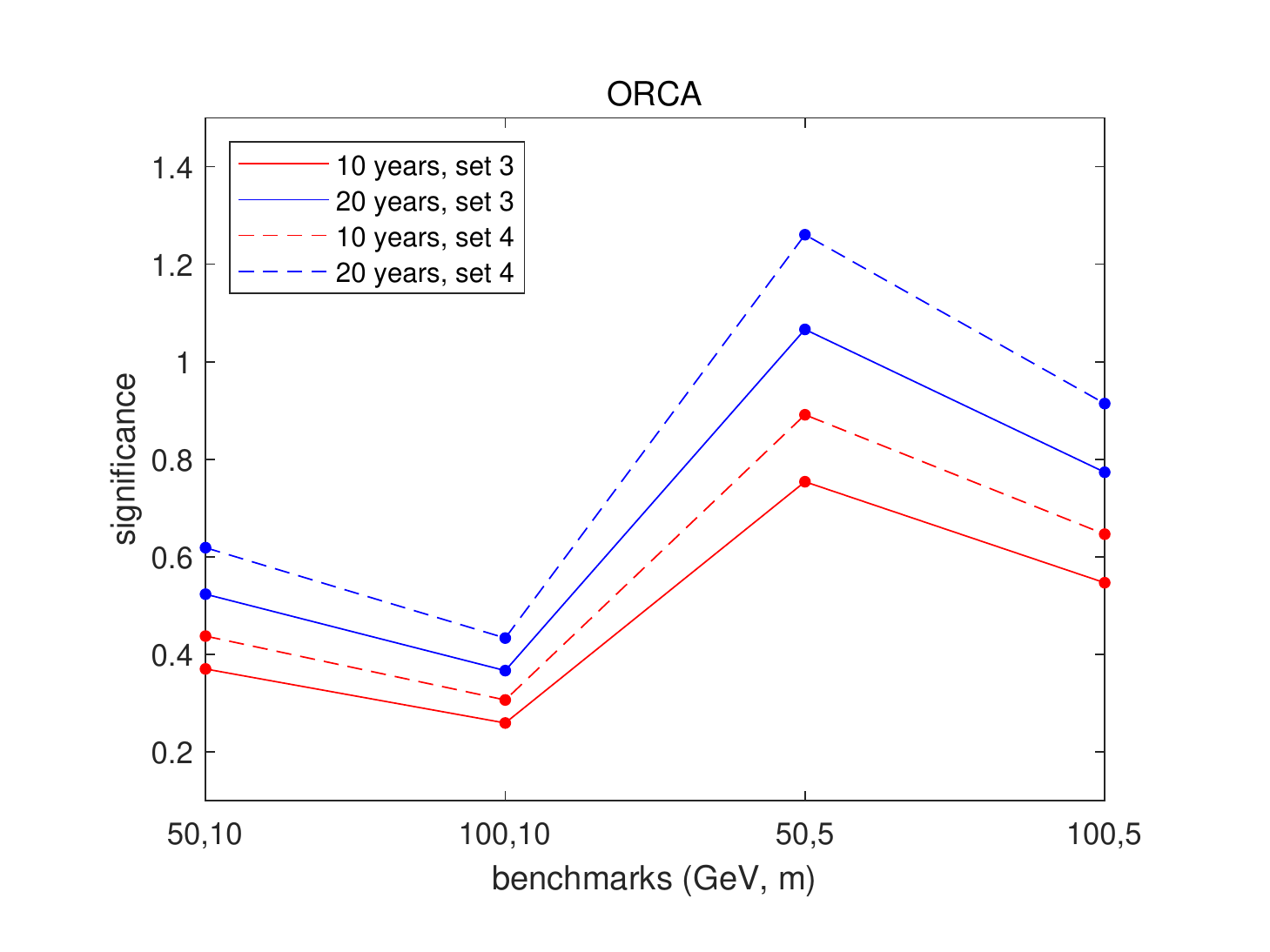}
    \includegraphics[width=0.49\textwidth]{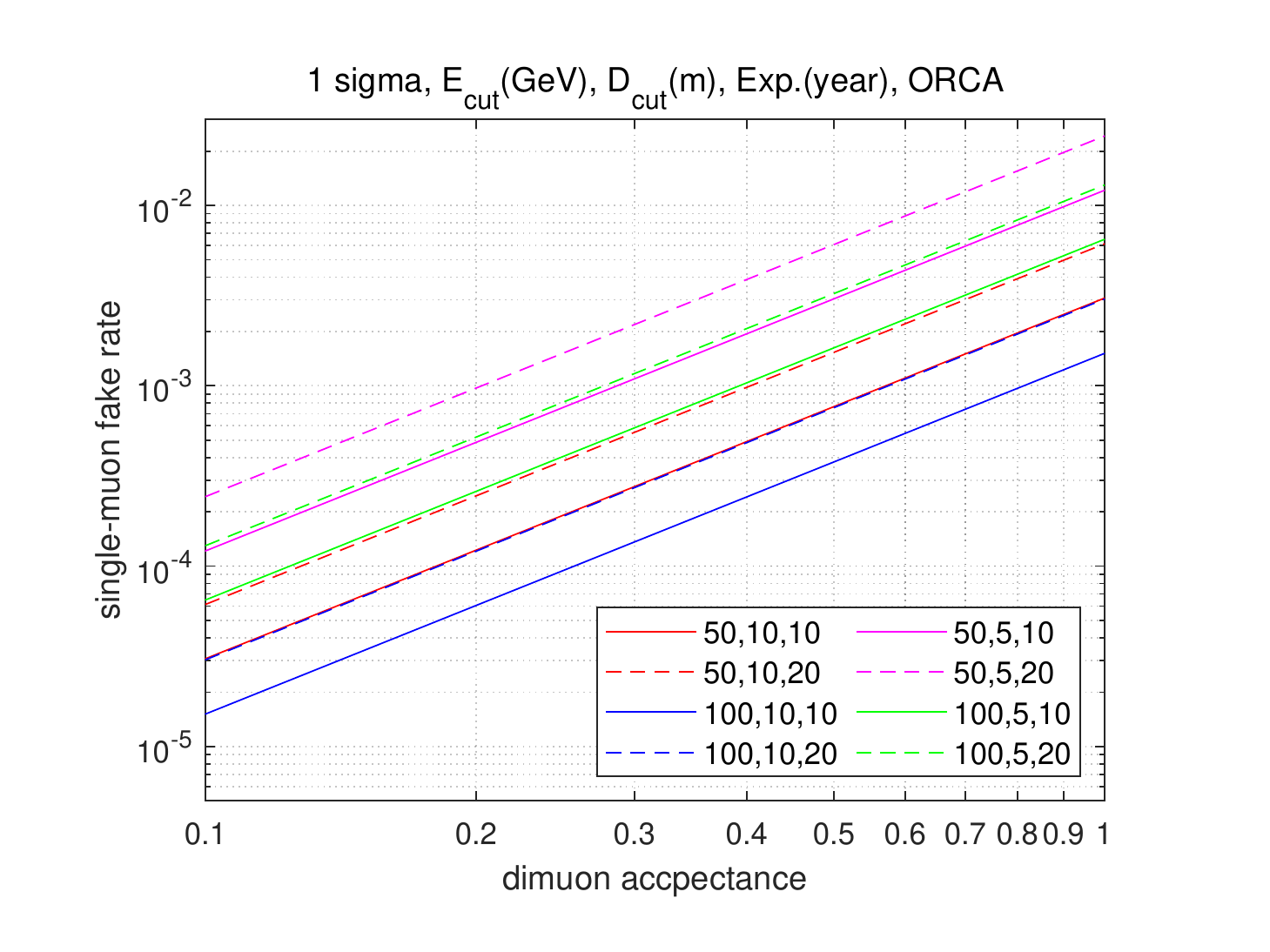}
    \caption{
    {\bf Upper-Left:} Significances of dimuon signal over single muon background are calculated with different detector-level classification (mis-)efficiencies (line and dashed line) for each theoretical benchmark (shown in x labels, e.g., (100,20) represents dimuon energy cut 100 GeV and dimuon separation cut 20 meters). Detector-level settings are based on IceCube configuration. Ten years (red) and twenty years (blue) exposures are taken.
    {\bf Upper-Right:} Critical dimuon acceptance and single-muon fake rate for achieving one sigma significance for different theoretical benchmarks (shown in legends, e.g., (50,5,10) represents dimuon energy cut 50 GeV, dimuon separation cut 5 meters and exposure time 10 years) in IceCube.
    {\bf Lower-Left:} The same as the upper-left panel, but detector-level settings are for ORCA. 
    {\bf Lower-Right:} The same as the upper-right panel, but modified for ORCA.
    }
    \label{fig:sig}
\end{figure*}
\subsection{Experimental Sensitivities}

Based on dimuon acceptances and single-muon fake rates computed in the Sec.~\ref{sec:eff}, experimental significances for several theoretical benchmarks are calculated.
Under background-only hypothesis, significance is calculated from $\text{sig} = \epsilon_{\mu\mu} N_{\mu\mu} / \sqrt{\epsilon_{\mu} N_{\mu}}$, here $N_{\mu\mu}$ and $N_{\mu}$ are event numbers of dimuons and single-muons that pass cuts, respectively.

This formula can be approximately factorized into two independent components: $\epsilon_{\mu\mu} / \sqrt{\epsilon_{\mu}}$ and $N_{\mu\mu} / \sqrt{N_{\mu}}$.
The former factor only relies on detector-level simulation results listed in Table~\ref{tab:efficiencies}, which are categorized into 4 detector-level sets.
The latter one is dependent on theoretical predictions.
By applying different cuts on muon energy and dimuon separation at the detector boundary, different events number of dimuon and single-muon are obtained and they constitute the theoretical benchmarks.  
In our estimation, the single-muon background is generated using the same flux and flux cut as that for dimuon events.
Under consideration of the energy information that can be used in a real experiment to enhance the power of classification, the energy cut at detector boundary for single-muon is twice the value of dimuon.
In Sec.~\ref{sec:comparison}, it is shown that the lateral distance of dimuon increases as they propagate inside the detector.
Hence, it is conservative to use dimuon separation as the counterpart of track separation mentioned in Sec.~\ref{sec:eff}.
In that subsection, dimuons in detector-level modelling are simulated with fixed separations, namely $30\,\text{m}$ and $10\,\text{m}$ for sparse and dense configuration, respectively. In contrast, dimuon events in theoretical benchmarks are considered with smaller separations.
For example, we consider dimuon separation cuts of either 30 meters or 20 meters in IceCube-like configuration.
In addition, for ORCA-like configuration, $N_{\mu\mu} / \sqrt{N_{\mu}}$ is rescaled by a factor 0.205 to account for the distinction of cross-sectional areas.
Results of our estimations are shown in the left panels of Figure~\ref{fig:sig}. The exposure time is taken either ten years or twenty years.
Note that, for IceCube-like configuration, no marked significances are achieved even for exposure of twenty years.
To enhance the significance, it is essential to lower the energy threshold and improve separation resolution.
This makes it more promising for ORCA to search dimuon events.
In the lower-left panel, experimental sensitivities of ORCA are displayed.
Significances achieved in ORCA are about one order higher than those in IceCube configuration.
For dimuon energy cut 50 GeV and separation cut 5 meters with exposure time twenty years, the sensitivity reaches its most optimistic value and surpasses significance of one sigma.
For these theoretical benchmarks, critical dimuon efficiencies and single-muon fake rates for achieving significance of one sigma are also shown in the right panels of Figure~\ref{fig:sig}.
In contrast with the simulation results in Table~\ref{tab:efficiencies}, the critical boundary lines are far-to-reach for IceCube.
Even for 100\% acceptance rate of dimuon signals, background rejections better than per-mille are required.
In comparison, the boundary lines for ORCA are lifted by orders, which is another manifestation of the advantages for ORCA to search for dimuon events.
We have not considered effects of various systematic uncertainties so far.
For instance, the flux uncertainties are about 20\% in the bulk region of neutrino energies
from theoretical models~\cite{Honda:2015fha}.
That can reduce the significance by a order of magnitude if adding directly to uncertainties
on predictions of background yields for configuration of either IceCube or ORCA.
However, one can use data driven methods to directly estimate the background yields from
measurements on normal single muon events~\cite{IceCube:2017cyo} without relying much on the theoretical inputs of neutrino fluxes.
That can also help with reducing impact of other experimental systematic uncertainties.
Besides, it must be emphasized that our phenomenological study is based on a preliminary reconstruction algorithm and simplified detector modeling. This is a starting point of future development. The optimization of hit splitting and the development of other sensitive selection criteria are promising to significantly improve the detecting potential. Furthermore, machine learning might be efficient to distinguish dimuon signals from single muon background.

\section{Conclusion}\label{sec:conclusion}
Dimuon events induced by charm-quark production in charged-current DIS merit investigation in neutrino telescopes.
For those with lateral separation of hundreds of meters, their discoveries would strongly imply physics beyond the Standard Model.
In the context of SM, dimuon events are still of great interests for the test of QCD.
However, short dimuon lateral distances make it challenging to find such signals in current neutrino telescopes.
In this paper, we set up a framework to simulate DIS dimuon events in two neutrino telescope models.
Only dimuon events produced outside the detector are considered.
Energy losses of muon traversing media are simulated with {\tt MMC}.
The resultant event sample from our simulation is exemplified in Figure~\ref{fig:ice_v} and Figure~\ref{fig:rock_v}.
Most events that can reach the detector occur near detector boundaries.
On the other hand, both the earlier work~\cite{Zhou:2021xuh} and our work discuss a similar scenario though developed independently. To compare with it in Sec.~\ref{sec:comparison},
simulation conditions are modified to match the baseline in that work.
In the first part of that section, the normalization of muons in ten years exposure are compared, 
and a further comparison of muon energy distributions shows consistency in patterns.
Furthermore, difference from calculation methods of muon energy losses is discussed.
To investigate the detection prospects in a real experiment, we set up a detection framework in Sec.~\ref{sec:analysis}. 
The efficiencies of four sets in two experimental configurations, namely IceCube-like and ORCA-like, are evaluated. With the help of these efficiencies, several theoretical benchmarks are selected to estimate the significance of dimuon signal over the single-muon background.
In this phenomenological study with over-simplified detector-level modelling, even the most optimistic significance achieved in IceCube is far less than one.
To achieve one sigma significance, the required dimuon acceptance rates and single-muon fake rates are also shown.
These critical lines are far-to-reach in the current simplified IceCube detection simulation.
It is therefore a big challenge to observe DIS dimuon signals in sparse km$^3$ scale neutrino telescopes.
By contrast, a denser but smaller configuration similar to ORCA, shows a better performance on dimuon searches. 
In the case of 50 GeV cut for energy and 5 meters cut for dimuon separation, it is expected to reach the significance level higher than one sigma in exposure of twenty years.
However, denser configurations are favored when the surrounding media can be used as targets. For detectors like Hyper-K, dimuon events can only be produced inside the detector, so challenge still remains for them.
In spite of limited sensitivities in this study, the dimuon search is still prospective. On the one hand, the detector-level simulation and reconstruction in this paper are simplified. By extending the detector response with PMT response modeling, time-correlated double pulse waveforms are expected on a series of DOMs.
Moreover, an optimization of double-track reconstruction with an additional small separation angle, instead of the parallel tracks assumed in the current algorithm, is helpful to improve the ability of classification. 
On the other hand, the detector design optimization would largely enhance the dimuon searches. In particular, the optimized optical sensors with much better timing resolution, for instance, a hybid-DOM consisting of multiple PMTs and SiPMs would significantly improve the direction resolution. This is potentially beneficial to the search of dimuons with short lateral distances. At present, such optimization is ongoing.

\begin{acknowledgments}
The authors would like to thank John Beacom and Bei Zhou for useful comments.
This work was sponsored by the National Natural Science Foundation of China under the Grant No.11875189 and No.11835005, and by Shanghai Jiao Tong University under the Double First Class Start-up Fund (WF220442603).
\end{acknowledgments}

\bibliography{references}

\end{document}